\documentclass[aps,prl,twocolumn,groupedaddress]{revtex4}
\usepackage{graphicx,amsfonts,color,epsfig}
\newcommand{\SU}[1]{\ensuremath{\mathrm{SU}( #1 )}}

\newcommand{\SpR}[1]{\ensuremath{\mathrm{Sp}( #1,\mathbb{R} )}}
\newcommand{\su}[1]{\ensuremath{\mathfrak{su}( #1 )}}

\newcommand{\so}[1]{\ensuremath{\mathfrak{so}( #1 )}}

\newcommand{\spR}[1]{\ensuremath{\mathfrak{sp}( #1, \mathbb{R} )}}
\newcommand{\etal}{\emph{et al.}}

\newcommand{\betb}{\begin{tabular}{p{4.0cm}p{9.0cm}}}
\newcommand{\entb}{\end{tabular}}

\newcommand{\ho}{\ensuremath{\hbar\Omega}}
\newcommand{\ph}[1]{\ensuremath{#1}p-\ensuremath{#1}h}
\begin{document}

\title{
Evidence for Symplectic Symmetry in {\it Ab Initio} \\  
No-Core Shell Model Results for Light Nuclei
}
\author{Tom\'a\v{s} Dytrych}
\author{Kristina D. Sviratcheva}
\author{Chairul Bahri}
\author{Jerry P. Draayer}
\affiliation{Department of Physics and Astronomy, Louisiana State
University, Baton
Rouge, LA 70803, USA}
\author{James P. Vary}
\affiliation{Department of Physics and Astronomy, Iowa State
University, Ames, IA 50011,
USA}
\affiliation{Lawrence Livermore National Laboratory, L-414, 7000 East
Avenue, Livermore,
California, 94551, USA}
\affiliation{Stanford Linear Accelerator Center, MS81, 2575 Sand Hill
Road, Menlo Park,
California, 94025, USA}

\begin{abstract}
Clear evidence for symplectic symmetry in low-lying states of $^{12}$C and $^{16}$O 
is reported. Eigenstates of $^{12}$C and $^{16}$O, determined within the framework of 
the no-core shell model using the JISP16 $NN$ realistic interaction, typically project 
at the 85-90\% level onto a few of the most deformed symplectic basis states that span 
only a small fraction of the full model space. The results are nearly independent of 
whether the bare or renormalized effective interactions are used in the analysis. The 
outcome confirms Elliott's \SU{3} model which underpins the symplectic scheme, and
above  all, points to the relevance of a symplectic no-core shell model
that can reproduce  experimental B(E2) values without effective charges as well as
deformed spatial modes associated with clustering phenomena in nuclei.  

\end{abstract}


\maketitle

Recently developed realistic interactions, such as $J$-matrix inverse scattering
potentials \cite{ShirokovMZVW04}  and modern two- and three-nucleon potentials derived
from meson exchange theory \cite{MachleidtSS96M01} or by using chiral  effective field
theory \cite{EntemM03},
succeed in modeling the essence of the strong interaction for
the purpose of input into microscopic shell-model calculations that
target reproducing characteristic features of light nuclei. The {\it ab initio} No-Core
Shell Model (NCSM) \cite{NCSM} which employs such modern realistic interactions, yields
a  good description of the low-lying states in few-nucleon
systems \cite{NavratilBK98_00} as well as in more complex nuclei like $^{12}$C
\cite{NCSM,NavratilO03}. In addition to advancing our understanding of the propagation
of the nucleon-nucleon force in nuclear matter and clustering phenomena
\cite{KarataglidisDAS95,FunakiTHSR03}, modeling the structure of $^{12}$C, $^{16}$O and
similar nuclei is also important for gaining a better understanding of other physical
processes such as parity-violating electron scattering from light nuclei
\cite{MusolfD92} and results gained through neutrino studies  \cite{HayesNV03} as well
as for making better predictions for capture reaction rates that figure prominently, for
example, in the burning of He in massive stars \cite{Brune99}.

In this letter we report on investigations that show that realistic eigenstates for
low-lying states determined in
NCSM calculations for
light nuclei with the JISP16 realistic interaction \cite{ShirokovMZVW04}, predominantly
project onto few of the most deformed \SpR{3}-symmetric basis states that are free of
spurious center-of-mass motion. This reflects the presence of an underlying symplectic
$\spR{3}\supset \su{3} \supset\so{3}$ algebraic structure \footnote{We use lowercase 
(capital) letters for algebras (groups).}, which is not {\it a priori} imposed on the
interaction and furthermore is found to remain unaltered after a  Lee-Suzuki similarity
transformation used to accommodate the truncation of the infinite Hilbert space by
renormalization of the bare interaction. This in turn provides insight into the physics
of a nucleon system and its geometry. Specifically, nuclear collective states  with
well-developed quadrupole and monopole vibrational modes and rotational modes are
described  naturally by
irreducible representations (irreps) of \SpR{3}. 

The present study points to the possibility of achieving convergence of higher-lying
collective modes and reaching heavier nuclei by expanding the NCSM basis space beyond
its current limits through \SpR{3} basis states that span a dramatically smaller
subspace of the full space. In this way, the symplectic no-core  shell-model (Sp-NCSM)
with realistic interactions and with a mixed \SpR{3} irrep extension will allow one to
account for even higher $\hbar\Omega$ configurations required to realize experimentally
measured B(E2) values without an effective charge, and 
to accommodate highly
deformed spatial configurations \cite{HechtB77Suzuki85} that are required to reproduce
$\alpha$-cluster modes, which may be responsible for shaping, e.g., the second $0^+$
state in $^{12}$C and $^{16}$O \cite{FunakiTHSR03}.

We focus on the $0^{+}_{gs}$ ground state and the lowest $2^{+}(\equiv\!2^+_1)$ and
$4^{+}(\equiv\!4^+_1)$ states in the oblate $^{12}$C nucleus as well as the $0^{+}_{gs}$
in the `closed-shell' $^{16}$O nucleus.  The NCSM eigenstates for these states
are reasonably well converged in the $N_{max}=6$ (or 6\ho) model space with an
effective interaction based on  the JISP16 realistic interaction \cite{ShirokovMZVW04},
which typically leads to rapid convergence in the NCSM evaluations, describes $NN$ data
to high accuracy and is consistent with,
but not constrained by, meson exchange theory, QCD or locality. In addition, calculated
binding energies as well as  other observables for $^{12}$C such as
B(E2;$2^{+}_{1}\!\rightarrow\!0^{+}_{gs}$), B(M1;$1^{+}_{1}\!\rightarrow\!0^{+}_{gs}$),
ground-state proton rms radii and the
$2^{+}_{1}$ quadrupole moment all lie reasonably close to the measured values.
While symplectic algebraic approaches have achieved a very good reproduction of
low-lying energies and B(E2) values in light nuclei \cite{RosensteelR80,DraayerWR84}
and specifically in $^{12}$C using phenomenological interactions \cite{EscherL02} or
truncated symplectic basis with simplistic (semi-) microscopic interactions
\cite{ArickxBD82,AvanciniP93}, here, for the first time, we establish, the dominance of
the symplectic \SpR{3} symmetry in light nuclei, and hence their propensity towards
development of collective motion, as unveiled through {\it ab initio} calculations of
the NCSM type starting with realistic two-nucleon interactions. 


The symplectic shell model \cite{Sp3R1,Sp3R2} is based on the noncompact symplectic
\spR{3} algebra. The classical realization of this symmetry underpins the dynamics of
rotating bodies and has been used, for example, to describe the rotation of deformed
stars and galaxies \cite{Rosensteel93}. In its quantal realization it is known to
underpin the successful Bohr-Mottelson collective model and has also been shown to
be a multiple oscillator shell generalization of Elliott's \SU{3} model. Consequently,
symplectic basis states bring forward important information about nuclear shapes and
deformation in terms of $(\lambda,\mu)$, which serve to
label the \SU{3} irreps within a given \SpR{3} irrep,
for example, $(0,0)$, $(\lambda,0)$ and $(0,\mu)$ describe spherical,
prolate and oblate shapes, respectively.

The significance of the symplectic symmetry for a microscopic description of a
quantum many-body system emerges from the physical relevance of its 21 generators
constructed as bilinear products of the momentum ($p_\alpha$) and coordinate ($q_\beta$)
operators, e.g.
$p_\alpha p_\beta$, $p_\alpha q_\beta$, and
$q_\alpha q_\beta$ with $\alpha$, $\beta$ = $x$, $y$, and $z$ for the
3 spatial directions. Hence, the many-particle kinetic energy, the mass quadrupole
moment operator, and the angular momentum are all elements of the
$\spR{3}\supset \su{3} \supset\so{3}$  algebraic structure. It also includes monopole
and quadrupole collective vibrations reaching beyond a single shell to 
higher-lying and core configurations, as well as vorticity degrees
of freedom for a description of the continuum from irrotational to rigid
rotor flows. Alternatively, the elements
of the \spR{3} algebra can be represented as bilinear products in harmonic oscillator
(HO) raising and lowering operators, which means the basis states of a \SpR{3} irrep
can be expanded in a 3-D HO ($m$-scheme) basis which is the same basis used in the NCSM,
thereby facilitating calculations and symmetry identification. 

The basis states within a \SpR{3} irrep are built by applying symplectic
raising operators to a \ph{n} ($n$-particle-$n$-hole, $n$ = 0, 2, 4, ...) lowest-weight
\SpR{3} state (symplectic bandhead), which is defined by the usual requirement that the
symplectic lowering operator annihilates it. The raising operator induces a 2\ho~ 1p-1h
monopole or quadrupole excitation (one particle raised by two shells) together with a
smaller 2\ho~ \ph{2} correction for eliminating the spurious center-of-mass motion.
If one were to include all possible lowest-weight
\ph{n} starting state configurations $(n
\leq N_{max})$, and allowed all multiples thereof, one would span the full
NCSM space.


The lowest-lying eigenstates of
$^{12}$C and
$^{16}$O were calculated using the NCSM as implemented through the
Many Fermion Dynamics ~(MFD) code \cite{Vary92_MFD} with an
effective interaction derived from the realistic JISP16 {\it NN}
potential \cite{ShirokovMZVW04} for different \ho~ oscillator strengths. 
For both nuclei we constructed all of the \ph{0} and 2\ho~\ph{2} (2 particles raised by one 
shell each) symplectic bandheads and
generated their \SpR{3} irreps up to $N_{max}=6$ (6\ho~ model space).
Analysis of overlaps of the symplectic states with the NCSM eigenstates
for 2$\hbar\Omega$, 4$\hbar\Omega$, and 6$\hbar\Omega$ model spaces ($N_{max}=2,4,6$) reveals the
dominance of the
\ph{0} \SpR{3} irreps.  For the $0^+_{gs}$ and the lowest $2^+$ and
$4^+$ states in $^{12}$C  there are nonnegligible overlaps for only 3 of the 13 \ph{0} \SpR{3} irreps,
namely, the leading (most deformed) representation specified by the shape deformation of its symplectic
bandhead, $(0~4)$, and carrying spin $S=0$ together with two $(1~2)$
$S=1$ irreps with different bandhead constructions for
protons and neutrons. For the ground state of $^{16}$O there is
only one possible \ph{0} \SpR{3} irrep, (0\,0) $S=0$. In addition, among the 2\ho~\ph{2} \SpR{3}
irreps only a small fraction contributes significantly to the overlaps and it includes the  
most deformed configurations that correspond to oblate shapes in $^{12}$C and prolate ones
in $^{16}$O.

\begin{figure}[t]
\includegraphics[width=0.45\textwidth]{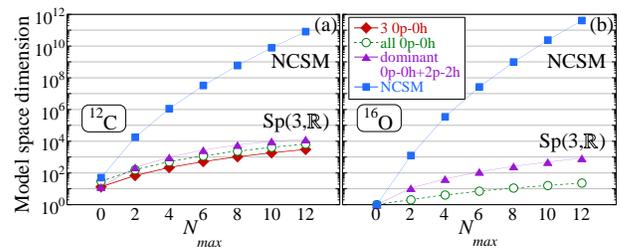}
\caption{
NCSM space dimension as a function of the maximum \ho~
excitations, $N_{max}$, compared to that of the \SpR{3} subspace: (a) $J=0,2,$ and
$4$ for
$^{12}$C, and (b) $J=0$ for
$^{16}$O.
}
\label{dimMdlSpace}
\end{figure}

The typical dimension of a symplectic irrep in
the $N_{max}=6$ space is on the order of $10^{2}$ as compared to
$10^{7}$ for the full NCSM $m$-scheme basis space.
As $N_{max}$ is increased the dimension of the $J=0,2,$ and
$4$ symplectic space  built on the \ph{0} \SpR{3} irreps for $^{12}$C
grows very slowly compared to the NCSM space dimension (Fig. \ref{dimMdlSpace}a). The
dominance of only three irreps additionally reduces the dimensionality of the symplectic
model space, which remains a small fraction of the NCSM basis
space even when the most dominant 2\ho~\ph{2} \SpR{3} irreps are included. The space reduction is even
more dramatic in the case of $^{16}$O (Fig.
\ref{dimMdlSpace}b). This means that a space spanned by a set of symplectic basis
states  is computationally manageable even when high-\ho~ configurations are included.

\begin{table}[b]
\caption{Probability distribution of NCSM eigenstates for $^{12}$C across the 
dominant \ph{0} and 2\ho~\ph{2} \SpR{3} irreps, \ho=15 MeV.\label{TABLE_15MeV}}
\begin{ruledtabular}
\begin{tabular}{lcrrrrr}
& & $0\hbar\Omega$ & $2\hbar\Omega$ & $4\hbar\Omega$ & $6\hbar\Omega$ & Total \\
\hline
\multicolumn{7}{c}{$J=0$} \\
\hline
$\SpR{3}$          & $(0\;4)S=0$     & $46.26$ & $12.58$ & $4.76$  & $1.24$  & $64.84$\\
                   & $(1\;2)S=1$     & $4.80$  & $2.02$  & $0.92$  & $0.38$  & $8.12$\\
   	               & $(1\;2)S=1$ & $4.72$  & $1.99$  & $0.91$  & $0.37$  & $7.99$ \\
            & 2\ho~\ph{2}            &         & $3.46$  & $1.02$  & $0.35$  & $4.83$\\
\cline{2-7}
                  & Total       & $55.78$& $20.05$& $7.61$ &  $2.34$  & $85.78$ \\
NCSM              &            & $56.18$ & $22.40$ & $12.81$ &  $7.00$  & $98.38$ \\
\hline
\multicolumn{7}{c}{$J=2$} \\
\hline
$\SpR{3}$          & $(0\;4)S=0$ & $46.80$ & $12.41$ & $4.55$  & $1.19$     &  $64.95$\\
                   & $(1\;2)S=1$ & $4.84$  & $1.77$  & $0.78$  &  $0.30$    &  $7.69$ \\
                   & $(1\;2)S=1$ & $4.69$  & $1.72$  & $0.76$  &  $0.30$    &  $7.47$ \\
                   & 2\ho~\ph{2} &         & $3.28$  & $1.04$  &  $0.38$    &  $4.70$\\
\cline{2-7}
     &    Total     & $56.33$ & $19.18$ & $7.13$ &  $2.17$  & $84.81$ \\
NCSM    &           & $56.18$ & $21.79$ & $12.73$ &  $7.28$ & $98.43$ \\
\hline
\multicolumn{7}{c}{$J=4$} \\
\hline
$\SpR{3}$          & $(0\;4)S=0$ & $51.45$ & $12.11$ & $4.18$   & $1.04$     & $68.78$\\
                   & $(1\;2)S=1$ & $3.04$  & $0.95$  & $0.40$   & $0.15$     & $4.54$ \\
                   & $(1\;2)S=1$ & $3.01$  & $0.94$   & $0.39$  & $0.15$     & $4.49$ \\
                  & 2\ho~\ph{2}  &         & $3.23$   & $1.16$  & $0.39$     & $4.78$\\
\cline{2-7}
       &    Total & $57.50$ & $17.23$ & $6.13$  &  $1.73$  & $82.59$ \\
NCSM   &          & $57.64$ & $20.34$ & $12.59$ &  $7.66$  & $98.23$
\end{tabular}
\end{ruledtabular}
\end{table}
\begin{table}[b]
\caption{Probability distribution of the NCSM eigenstate for the $J=0$ ground state in
$^{16}$O across the \ph{0} and dominant 2\ho~\ph{2} \SpR{3} irreps, \ho=15 MeV.
\label{TABLE_O16}}
\begin{ruledtabular}
\begin{tabular}{lcrrrrr}
& & $0\hbar\Omega$ & $2\hbar\Omega$ & $4\hbar\Omega$ & $6\hbar\Omega$ & Total \\
\hline
$\SpR{3}$ & $(0\;0)S=0$ & $50.53$ & $15.87$ & $6.32$ & $2.30$ & $75.02$ \\
          & 2\ho~\ph{2} & & $5.99$ & $2.52$ & $1.32$ & $9.83$ \\
\cline{2-7}
                 & Total       & $50.53$ & $21.86$  & $8.84$  &  $3.62$ & $84.85$ \\
NCSM &                         & $50.53$ & $22.58$ & $14.91$ &  $10.81$ & $98.83$ \\
\end{tabular}
\end{ruledtabular}
\end{table}
The overlaps of the most dominant symplectic states with investigated NCSM eigenstates for the 
$^{12}$C and the $^{16}$O in the $0$, $2$, $4$ and $6\hbar\Omega$ subspaces are 
given in Table~\ref{TABLE_15MeV} and \ref{TABLE_O16}.
In order to speed up the calculations, we retained only the largest amplitudes of the NCSM states,
those sufficient to account for at least 98\% of the norm which is quoted also in the table. 
The results show that approximately 85\% of the NCSM eigenstates for $^{12}$C
($^{16}$O) fall within a subspace spanned by the few most significant \ph{0} and 2\ho~\ph{2} \SpR{3}
irreps, with the 2\ho~\ph{2} \SpR{3} irreps accounting for 5\% (10\%) and with the leading irrep,
$(0~4)$ for $^{12}$C and $(0~0)$ for $^{16}$O, carrying close to 70\% (75\%) of the NCSM
wavefunction.

Furthermore, the $S=0$ part of all three  NCSM eigenstates for $^{12}$C is almost
entirely projected (95\%) onto only six $S=0$ symplectic irreps included in Table
\ref{TABLE_15MeV}, with as much as 90\% of the spin-zero NCSM states accounted for
solely by  the leading $(0\,4)$ irrep. 
The $S=1$ part is also remarkably well described by merely two \SpR{3} irreps. 
Similar results are observed for the ground state of $^{16}$O.

\begin{figure}[th]
\centerline{\epsfxsize=3.4in\epsfbox{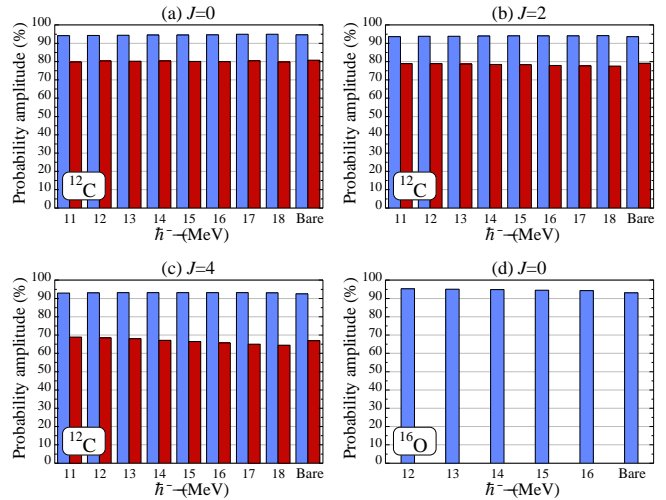}} 
\caption{
Projection of the $S=0$ (blue, left) [and $S=1$ (red, right)] \SpR{3} irreps onto the
corresponding significant spin components of the NSCM wavefunctions for (a) $0^{+}_{gs}$, (b)
$2^{+}_1$, and (c)
$4^{+}_1$ in
$^{12}$C and (d) $0^{+}_{gs}$ in $^{16}$O, for effective interaction for different
$\hbar\Omega$ oscillator strengths and bare interaction.
}
\label{Overlaps_SpinScaled}
\end{figure}

Another striking property of the low-lying eigenstates is revealed when the spin
projections of the converged NCSM states are examined. Specifically, as shown in
Fig.~\ref{Overlaps_SpinScaled}, their \SpR{3} symmetry and hence the geometry of the
nucleon system being described is nearly independent of the \ho~oscillator strength.
The symplectic  symmetry is present with equal strength in the spin parts of the NCSM
wavefunctions for $^{12}$C as well as $^{16}$O regardless of whether the bare or the
effective interactions are used. This suggests that the  Lee-Suzuki transformation,
which effectively compensates for the finite space truncation by  renormalization of
the bare interaction, does not affect the \SpR{3} symmetry structure of the spatial
wavefunctions. Hence, the symplectic structure detected in the present analysis for
$6\ho$ model space is what would emerge in NSCM evaluations with a sufficiently large
model space to justify use of the bare interaction.

\begin{figure}[th]
\centerline{\epsfxsize=2.9in\epsfbox{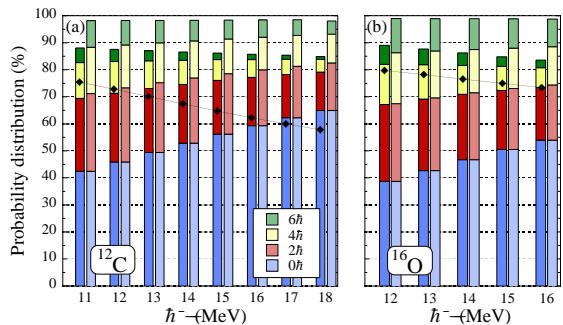}}
\caption{
Ground $0^+$ state probability distribution 
over $0$\ho~ (blue, lowest) to $6$\ho~ (green, highest) subspaces
for the most dominant \ph{0} + 2\ho~\ph{2} \SpR{3} irrep case 
(left) and NCSM (right) together with the leading irrep
contribution (black diamonds), $(0~4)$ for $^{12}$C (a) and $(0~0)$ for $^{16}$O (b), as a function of
the
\ho~ oscillator strength, $N_{max}=6$.}
\label{C12prblty_vs_hw}
\end{figure}

In addition, as one varies the oscillator strength \ho, the projection of the NCSM
wavefunctions onto the symplectic subspace changes only slightly (see, e.g.,
Fig. \ref{C12prblty_vs_hw} for the $0^+_{gs}$ state of $^{12}$C and $^{16}$O). The
symplectic structure is preserved, only the \SpR{3} irrep contributions change because
the $S=0$ ($S=1$) part of the NCSM eigenstates decrease (increase) towards higher \ho~
frequencies.  Clearly, the largest contribution comes from the leading
\SpR{3} irrep (black diamonds), growing to 80\% of the NCSM wavefunctions for the
lowest \ho.
These results can be interpreted as a strong confirmation of Elliott's \SU{3} model
since the projection of the NCSM states onto the 0$\hbar\Omega$ space [Fig.
\ref{C12prblty_vs_hw}, blue (lowest) bars] is a projection of the NCSM results onto
the \SU{3} shell model. The outcome is consistent with what has been shown to be a
dominance of the leading \SU{3} symmetry for \SU{3}-based shell-model studies with
realistic interactions in 0$\hbar\Omega$ model spaces.  It seems the simplest of
Elliott's collective states can be regarded as a good first-order approximation in the
presence of realistic interactions, whether the latter is restricted to a
0$\hbar\Omega$ model space or richer multi-$\hbar\Omega$ NCSM model
spaces.

The $0^{+}_{gs}$ and $2^{+}_1$ states in $^{12}$C, constructed in terms of the
three \SpR{3} irreps with probability  amplitudes defined by the
overlaps with the NCSM wavefunctions for $N_{max}=6$ case, were also used to determine
$B(E2:2^+_1\rightarrow0^+_{gs})$ transition rates. The latter, increasing
from $101\%$ to $107\%$ of the corresponding NCSM numbers with increasing \ho,
clearly reproduce the NCSM results. 

In summary, we have shown that {\it ab initio} NCSM calculations with the JISP16
nucleon-nucleon interaction display a very clear symplectic structure, which
is unaltered whether the bare or effective interactions for various \ho~
strengths are used. Specifically, NCSM wavefunctions for the lowest
$0^+_{gs}$, $2^+_1$ and $4^+_1$ states
in $^{12}$C and the ground state in $^{16}$O
project at the 85-90\% level onto a few \ph{0} and 2\ho~\ph{2}
spurious center-of-mass free symplectic irreps.
Furthermore, while the dimensionality of the latter is only
$\approx 10^{-3}\%$ that of the NCSM space, they closely reproduce the NCSM $B(E2)$
estimates. The wavefunctions for $^{12}$C are
strongly dominated by the three leading
\ph{0} symplectic irreps, with a clear dominance of
the most deformed $(0\,4)S=0$ collective configuration. The ground state of $^{16}$O is 
dominated by the single \ph{0} irrep $(0\,0)S=0$. 
The results confirm for the first time the validity of the \SpR{3} approach when
realistic interactions are invoked in a NCSM space. This demonstrates the importance of
the \SpR{3} symmetry in light nuclei while reaffirming the value of the simpler
\SU{3} model upon which it is based. The results further suggest that a Sp-NCSM
extension of the NCSM may be a practical scheme for achieving convergence to measured
$B(E2)$ values without the need for introducing an effective charge. In short, the
NCSM with a modern realistic interaction supports the development of collective
motion in nuclei which is realized through the Sp-NCSM and as is
apparent in its 0$\hbar\Omega$ Elliott model limit. 


Discussions with many colleagues, but especially Bruce R. Barrett, are gratefully
acknowledged.  This work was supported by the US National Science Foundation, Grant
Nos 0140300 \& 0500291, and the Southeastern Universities Research Association, as
well as, in part, by the US Department of Energy Grant Nos.  DE-AC02-76SF00515 and
DE-FG02-87ER40371 and at the University of California, Lawrence Livermore National
Laboratory under contract No. W-7405-Eng-48. Tom\'{a}\v{s} Dytrych acknowledges supplemental support from
the Graduate School of Louisiana State University.

\end{document}